# Giant Magnetoimpedance in Crystalline Mumetal

H. B. Nie[1], A. B. Pakhomov[1,2], X. Yan[1], X. X. Zhang[1] and M. Knobel[2]

[1]*Department of Physics, Hong Kong University of Science and Technology, Clear Water Bay, Kowloon, Hong Kong, People's Republic of China*

[2]*Instituto de Física "Gleb Wataghin", Universidade Estadual de Campinas (UNICAMP), C.P. 6165, Campinas 13083-970, S.P., Brazil.*

**Abstract**

We studied giant magnetoimpedance (GMI) in commercial crystalline Mumetal, with the emphasis to sample thickness dependence and annealing effects. By using appropriate heat treatment one can achieve GMI ratios as high as 310%, and field sensitivity of about 20%/Oe, which is comparable to the best GMI characteristics obtained for amorphous and nanocrystalline soft magnetic materials.

Giant magnetoimpedance (GMI) is the name given to the considerable change of the complex impedance when soft ferromagnetic conducting samples are subjected to an external magnetic field [1]. GMI has been studied in amorphous and nanocrystalline wires [1-6], ribbons [7-8], films [9-12], single crystals [13] and fibers [14] with high magnetic permeability. The high sensitivity of the GMI to the applied dc field has been shown useful for potential applications in magnetic sensing elements [1, 10, 12, 15]. The GMI effect is due to a dependence of the transverse magnetic permeability $\mu_t$ on the applied longitudinal magnetic field *H*. At relatively low frequency, this dependence will affect only the reactive component of the impedance through the sample inductance, and the effect is known as magnetoinductance [10-12]. GMI is observed at higher frequencies when the magnetic penetration depth, or skin depth [16]

$$\delta = \sqrt{\frac{\rho}{\pi f \mu_0 \mu_t}}, \qquad (1)$$

is smaller than the sample thickness [1, 2, 6]. Here $\rho$ is the resistivity, *f* the frequency and $\mu_0$ the magnetic permeability of vacuum. In this case, the applied current flows effectively just in a shell of thickness $\delta$ near the surface [16], giving rise to increased values of both the real and imaginary parts of the impedance. Application of the axial magnetic field leads to a considerable decrease of the transverse permeability, increase of the penetration depth, and hence to a decrease of both the resistance and the reactance.

Although GMI has been widely studied before in many amorphous and nanocrystalline systems, which are sometimes produced by quite complicated techniques, very little work has been reported on GMI in more conventional and inexpensive crystalline soft magnetic materials [17]. In this work, we study the GMI effect in stripe-shaped crystalline Mumetal. Mumetals are commercially available soft magnetic alloys, having relatively stable crystalline structure with respect to thermal or mechanical treatments and good thermal conductivity. The use of such materials for magnetic sensing may be therefore beneficial, if high GMI sensitivities are achieved.

GMI was studied both in the as-cast and annealed samples of varying thickness. We show that by appropriate heat treatment of Mumetal one can obtain a material highly sensitive to magnetic field. After annealing at optimum conditions, the maximum GMI ratio $(\frac{\Delta Z}{Z})_{max}$ increased by a factor of 3 (from 90 % to about 310 % at 600 kHz), and the maximum sensitivity $\left| [d(\frac{\Delta Z}{Z})/dH] \right|_{max}$ increased by a factor of 10 (from 1.8 %/Oe to about 21 %/Oe at 600 kHz). The value of $(\frac{\Delta Z}{Z})_{max}$ = 310% obtained in our experiments is a factor of 2 higher than that previously reported for NiFeMo permalloy wires [17]. The frequency and thickness dependencies of the giant magneto-impedance (GMI) can be well explained by the classical electrodynamical model.

The measurements were done on stripe-shaped samples cut from crystalline Mumetal (Goodfellow Cambridge Limited) foils of varying thickness, 1.0 mm, 125 µm, 50 µm, 25 µm and 12.5 µm. Samples had lengths between 15-25 mm, and widths in the range 0.5-1.0 mm. The nominal composition of Mumetal was $Ni_{77}Fe_{14}Cu_5Mo_4$. Annealing was carried out in the range $T_a$ = 430 - 800 °C, annealing time varying between 20 min and 3 h.

HP 4284A and HP 4285A LCR Meters were used to measure the impedance Z, resistance R and reactance X, where $Z = |R + iX|$ with a four-probe technique. The measurements were done at the rms current magnitude $I \approx 10$ mA. A Helmholtz coil was used to generate a magnetic field parallel to the stripe axis. The data were obtained from two types of quasi-static scans. Magnetic field scans were performed in the range -115.5 Oe $\leq H \leq$ 115.5 Oe at a fixed frequency. Frequency scans covered the range 20 Hz $\leq f \leq$ 1 MHz or 75 kHz $\leq f \leq$ 30 MHz with either H=0, or H=115.5 Oe, the latter value being above the magnetoimpedance saturation. We define the GMI ratio as $\Delta Z/Z = [Z(0)-Z(H_s)]/Z(H_s)$, where $H_s$ is the magnetic field where the value of impedance saturates. The sensitivity of the impedance response is defined as a maximum derivative $\left| [d(\frac{\Delta Z}{Z})/dH] \right|_{max}$.

A typical example of frequency dependence of impedance is shown on a double logarithmic scale in Fig. 1. Here the frequency dependencies of the zero field resistance $R_0$ and reactance $X_0$, and the saturated values of resistance $R_s$ and reactance $X_s$ (the latter values measured at $H = 115.5$ Oe) are shown for a sample with the thickness $h = 125$ μm after annealing for 40 minutes at 580 °C. The saturated resistance $R_s$ is a constant, and the saturated reactance $X_s \equiv 2\pi f \times L_s$ is a linear function of frequency (constant inductance $L_s$), in the whole measurement range. This is apparently because at saturation the current density is almost homogeneous in the sample cross section.

The behavior of the zero-field impedance (Fig. 1) is different. At relatively low frequencies ($f < 20$ kHz) both the resistance $R_0$ and inductance $L_0 \equiv X_0/2\pi f$ are constant, but the inductance value is considerably greater than at saturation, reflecting the difference in magnetic permeability of the sample. For a strip sample of the length $l$, width $w$ and thickness $h$, where $w \gg h$, with the uniform current distribution, the resistance and reactance can be expressed as follows:

$$R = \rho \frac{l}{wh} \text{, and } X \equiv 2\pi f L = \frac{\mu_0 \mu h \pi f}{4w} l \quad (2)$$

From the expression for $X$ and the low frequency data of Fig. 1, it follows that the magnetic permeability drops by a factor of over 20 between the zero-field and saturated states. The value of the transverse permeability can also be approximately estimated from this data. Using the expression for the reactance (Eq. 2) we find that $\mu(H = 0) \sim 10^3$.

The log-log frequency dependencies of both the resistance and reactance in zero field change at high frequencies (above approximately 20 KHz), where the slopes of both curves for $R_0$ and $X_0$ in Fig. 1 cross over to 0.5. In other words, at high frequencies both the zero field resistance $R_0$ and reactance $X_0$ are proportional to $f^{1/2}$. Obviously this is because, for large values of permeability, the skin depth $\delta_m$ defined by Eq. (1) becomes smaller than the sample thickness. Substitution of the thickness $h$ in Eq. (2) with the penetration depth gives the square root dependences for both $R_0$ and $X_0$.

Fig. 2 shows the magnetoimpedance ratio $\frac{\Delta Z}{Z}$ versus frequency for as-cast Mumetal samples of varying thickness $h$. For relatively low frequencies, the impedance $Z$ is almost independent on magnetic field, since the skin depth $\delta$ is larger than half of the thickness $h$, and the field-dependent $X$ is much smaller than the field-independent $R$. At the onset frequency $f_o$ (when $\delta \approx h/2$), the application of an external dc field begins to reduce impedance $Z$ as a consequence of the change in $\delta$. The onset frequency, which corresponds to the crossover in the $X_0$ and $R_0$ behavior in Fig. 1, can be obtained from the condition $\delta = h/2$. Therefore

$$f_o = \frac{4\rho}{\pi \mu_0 \mu_t^0 h^2}, \quad (3)$$

where $\mu_t^0$ is the relative transverse magnetic permeability when the external field $H= 0$. The onset frequency is proportional to $h^{-2}$. This is why a rapid increase of $\frac{\Delta Z}{Z}$ starts at a lower frequency for thicker samples as seen in Fig. 2.

With frequency increasing over $f_o$, the magnetoimpedance ratio increases and reaches a maximum at the peak frequency $f_p$. The existence of the maximum can be qualitatively understood from Fig. 1. The zero-field impedance is increasing at $f > f_0$ approximately as $f^{1/2}$, while the saturated value of $Z$ remains weakly-dependent on frequency until $X_s$ reaches the value of $R_s$. As the saturated permeability is small, the value of $Z_s$ continues to increase linearly with frequency, that is faster than $Z_0$, providing the maximum in $\frac{\Delta Z}{Z}$. Using Eq. (2), the peak frequency can be roughly estimated from the condition $R_s=X_s$ as

$$f_p = \frac{4\rho}{\pi\mu_0\mu_t^s h^2}, \qquad (4)$$

where $\mu_t^s$ is the saturated transverse permeability. Notice that in this approximation the onset frequency (3) and the peak frequency (4) have similar thickness dependences, and differ only by a factor equal to the ratio of the zero-field and saturated values of transverse permeability, or approximately by a factor of 20. At even higher frequencies $\frac{\Delta Z}{Z}$ will not become zero as the saturated impedance will eventually also cross over to a square root dependence on frequency. One can also show that the maximum MI ratio $(\frac{\Delta Z}{Z})_{max}$ is an increasing function of both the thickness and the ratio of permeabilities.

Fig. 3 shows the MI ratio $\frac{\Delta Z}{Z}$ versus frequency $f$ for a Mumetal sample of thickness $h = 125$ μm after annealing at different conditions: (a) with the same annealing time $t_a = 40$ minutes at different temperatures $T_a$, and (b) at the same annealing temperature $T_a = 580$ °C for different time $t_a$. The largest MI ratio was obtained after annealing for 40 minutes at $T_a=580$ °C. In this case, when the driving current frequency $f = 600$ KHz, the maximum MI ratio is about 310%, and the maximum sensitivity $\left|[d(\frac{\Delta Z}{Z})/dH]\right|_{max}$ is about 21%/Oe. These values are to be compared with $(\frac{\Delta Z}{Z})_{max} = 90$ %, and $\left|[d(\frac{\Delta Z}{Z})/dH]\right|_{max} = 1.8$ %/Oe for the as-cast sample.

From Fig. 3 (a), one can see that the value of the maximum GMI ratio and the onset frequency $f_o$ are sensitive to the annealing conditions, while the peak frequency $f_p$ almost does not change with anneling, at least when $T_a \leq 700$ °C. Using Eqs. (3), (4) we conclude that the electrical resistivity $\rho$ and the saturated magnetic permeability $\mu_t^s$ do not change considerably during annealing, while the zero field permeability is highly sensitive to heat treatment.

Enhanced GMI effects after annealing in a furnace or by electric current have been observed before in amorphous materials [4, 5, 11, 17-19]. Usually the enhancement of GMI after annealing was attributed to a further decrease of small negative magnetostriction and achieving softer magnetic properties. Heat treatment has long been recognized as an important tool for improvement of magnetic properties of nickel-iron alloys and their ternary and quaternary derivatives such as Mumetal [20]. Generally, the rotation part of magnetic susceptibility increases with decreasing of both magnetocrystalline anisotropy $K_1$, the magnetostriction $\lambda_s$ and the level of stresses in the material. We suggest that annealing at moderate conditions, such as 40 minutes at 580 $^0$C, affects permeability mainly via relaxation of mechanical stresses such as dislocation creep. That results in de-pinning of domain walls and increase of rotational part of permeability. A decrease of $\lambda_s$ with heat treatment has also been reported in the literature [20]. While further annealing at longer times or higher temperatures may lead to change of the phase composition of the solid solution, causing a deterioration of soft magnetic properties. For NiFe alloys it can be associated with ordering which increases magnetocrystalline anisotropy. We used optical microscopy to study the possible structural changes in polished sections of the annealed samples. No apparent changes in the grain geometry were observed upon annealing for 40 minutes up to the temperatures $T_a$ = 700 $^0$C, however we saw definite recrystallization at $T_a$ ~ 800 $^0$C and above. We assume that some phase changes, though indistinguishable by optical means, might have happened at lower temperature as well. However, in order to singe out the primary physical cause for the observed phenomena more magnetic studies must be performed.

To conclude, we presented a study of GMI for crystalline Mumetal in the range 20 Hz - 30 MHz. Vacuum annealing of a 125-micron-thick sample, at optimum conditions, leads to a considerable increase of both the maximum GMI ratio up to 310% (for maximum fields of the order of 100 Oe), the field sensitivity up to 21%/Oe. These values are comparable to the best GMI properties already observed in any soft magnetic material. We suggest that the effect of annealing can be owing mainly to the removal of stresses that develop in the process of the metal sheet manufacturing. However, further annealing leads to a decrease of the effect, supposedly due to phase changes in the material. We conclude that the achieved high sensitivity of the effect, combined with relative thermal stability, can make Mumetals candidate materials for applications in field sensing elements[1]. The GMI effect and its relation to magnetic permeability are well described by the classical Electrodynamics model, which considers the dependence of the impedance on the magnetic field due to the field-dependent transverse magnetic permeability.

**Acknowledgments**

The authors would like to acknowledge X. N. Jing, G. G. Zheng, Kingsley K. L. Wong, X. F. Zhang and J. B. Gen for technical assistance. This work was supported by HKUST6129/97P (Hong Kong). Also, Brazilian agencies FAPESP and CNPq are acknowledged for their financial support.

**References**

[1] M. Knobel, J. Phys. IV (France) PR2 (1998) 213.
[2] L.V. Panina, K. Mohri, Appl. Phys. Lett. vol. 65 (1994) 1189.
[3] J. Velázquez, M. Vázquez, D. X. Chen, A. Hernando, Phys. Rev. B 5 (1994) 16737.
[4] J.L. Costa-Krämer, K.V. Rao, IEEE Trans. Magn. 31 (1995) 1261.
[5] R. Valenzuela, M. Knobel, M. Vázquez, A. Hernando, J. Phys. D: Appl. Phys. 28 (1995) 2404.
[6] R.S.Beach, A.E. Berkowitz, Appl. Phys. Lett. 64 (1994) 3652.
[7] R. L. Sommer, C. L. Chien, Appl. Phys. Lett. 67 (1995) 857.
[8] F.L.A. Machado, C.S. Martins, S.M. Rezende, Phys. Rev. B 51 (1995) 3926.
[9] R. L. Sommer, C. L. Chien, Appl. Phys. Lett. 67 (1995) 3346.
[10] L.V. Panina, K.Mohri, T.Uchiyama, M.Noda, IEEE Trans. Magn. 31 (1995) 1249.
[11] T. Uchiyama, K. Mohri, L. V. Panina, K. Furuno, T. IEEE Jpn 115-A (10) (1995) 949.
[12] K. Hika, L.V. Panina, K. Mohri, IEEE Trans. Magn. 32 (1996) 4594.
[13] M. Carara, R. L. Sommer, J. Appl. Phys. 81 (1997) 4107.
[14] D. Menard, D, M. Britel, P. Ciureanu, A. Yelon, V.P. Paramonov, A.S. Antonov, P. Rudkowski, J.O. Strom-Olsen, J. Appl. Phys. 81 (1997) 4032.
[15] U. Barjenbruch, Sensors and Actuators A 37/38 (1993) 466.
[16] L. D. Landau, E. M. Lifschitz, Electrodynamics of Continuous Media, 2, Pergamon Press, New York, 1984.
[17] M. Vázquez, J. M. García-Beneytez, J. P. Sinneker, J. Appl. Phys. 83 (1998) 6578.
[18] Y. Takemura, H. Tokuda, K. Komatsu, S. Masuda, T. Yamada, K. Kakuno, K. Saito, IEEE Trans. Magn. 32 (1996) 4947.
[19] H.B. Nie, X.X. Zhang, A.B. Pakhomov, Z. Xie, X. Yan, A. Zhukov, M. Vazquez, J. Appl. Phys. 85 (1999) 4445.
[20] R. S. Tebble, D. J. Craik, Magnetic Materials, Wiley, London, 1969.

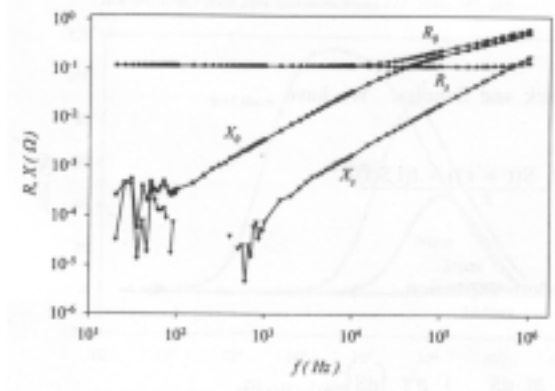

**Fig. 1.** Frequency dependencies (20 Hz ≤ f ≤1 MHz ) of resistance and reactance for zero applied field ($R_0$ and $X_0$) and at H= 115.5 Oe ($R_s$ and $X_s$) for a sample of thickness $h = 125$ μm after 40 min annealing at 580 °C.

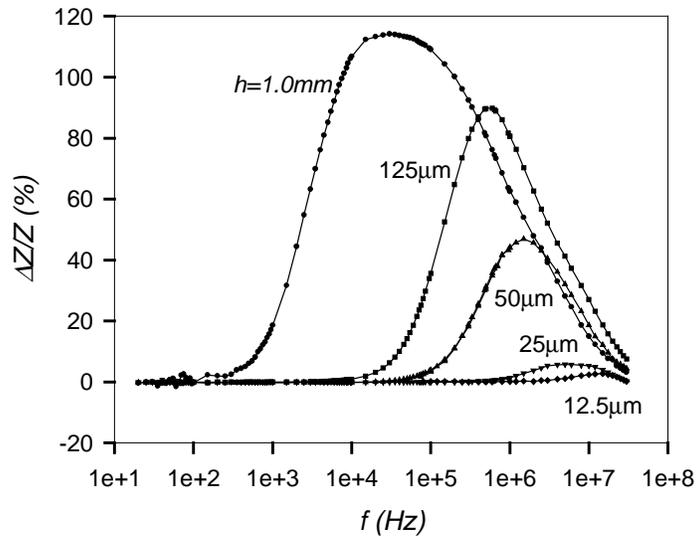

**Fig. 2.** Frequency dependencies (20 Hz ≤ $f$ ≤ 30 MHz) of GMI ratio $\dfrac{\Delta Z}{Z}$ for as-cast samples of varying thickness $h$.

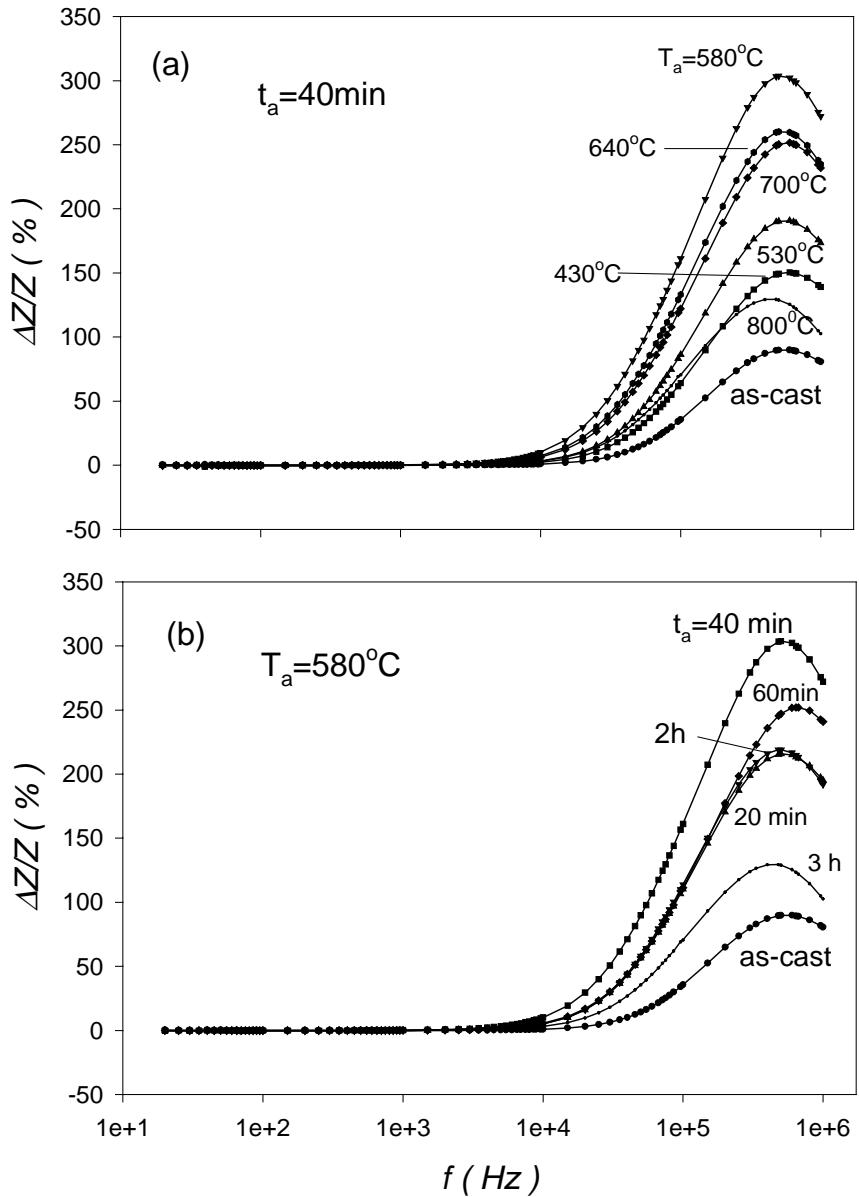

**Fig. 3** MI ratio $\dfrac{\Delta Z}{Z}$ versus frequency *f* for a 125 μm thick sample after annealing at varying conditions: (a) at constant annealing time $t_a$ = 40 min and increasing temperature; (b) at constant temperature $T_a$ = 580 °C with increasing annealing time.